\documentclass[pra,preprint]{revtex4}
\usepackage{epsfig,amssymb,amsfonts}
\begin{document}
%\doi{10.1080/0950034YYxxxxxxxx}
% \issn{1362-3044}
%\issnp{0950-0340} %\jvol{00}
%% \jnum{00} \jyear{2005} \jmonth{10 January}

%
%\markboth{Taylor \& Francis}{Transfer of a Polaritonic Qubit through
%a Coupled Cavity Array}

\title{ Transfer of a Polaritonic Qubit through a Coupled Cavity
Array}

\author{Sougato Bose $^{1}$}
\author{Dimitris G. \surname{Angelakis}$^{2}$}\email{dimitris.angelakis@qubit.org}
\author{Daniel  \surname{Burgarth}$^{1,3}$}%
  \address{$^{1}$Department of Physics and Astronomy, University
College London, Gower St., London WC1E 6BT, UK}
\address{$^{2}$Centre for Quantum Computation, Department
of Applied Mathematics
 and Theoretical Physics, University of Cambridge,
 Wilberforce Road, CB3 0WA, UK}
\address{$^{3}$Computer Science Departement,
ETH Z{\"u}rich, CH-8092 Z{\"u}rich, Switzerland\\\vspace{6pt}\received{7 Feb, 2007}}

\begin{abstract}
We demonstrate a scheme for quantum communication between the ends
of an array of coupled cavities.  Each cavity is doped with a single
two level system (atoms or quantum dots) and the detuning of the
atomic level spacing and photonic frequency is appropriately tuned
to achieve photon blockade in the array. We show that in such a
regime, the array can simulate a dual rail quantum state transfer
protocol where the arrival of quantum information at the receiving
cavity is heralded through a fluorescence measurement. Communication
is also possible between any pair of cavities of a network of
connected cavities.\bigskip

\end{abstract}
\maketitle

\section{Introduction}

Recently, the exciting possibility of coupling high Q cavities
directly with each other has materialized in a variety of settings,
namely fiber coupled micro-toroidal cavities \cite{armani-vahala03},
arrays of defects in photonic band gap materials (PBGs)
\cite{pbg,angelakis-knight04} and microwave stripline resonators
joined to each other \cite{supercond}. A further exciting
development has been the ability to couple each such cavity to a
quantum two-level system which could be atoms for micro-toroid
cavities, quantum dots for defects in PBGs or superconducting qubits
for microwave stripline resonators\cite{coupled_twolevel}.
Possibilities with such systems are enormous and include the the
implementation optical quantum computing \cite{angelakis-ekert04},
the production of entangled photons \cite{angelakis-bose06}, the
realization of Mott insulating and superfluid phases and spin chain
systems
\cite{angelakis-bose06b,hartmann-plenio06,greentree-hollenberg06} .
Such settings can also be used to verify the possibilities of
distributed quantum computation involving atoms coupled to distinct
cavities \cite{alessio} also to generate cluster states for
efficient measurement based quantum computing
schemes\cite{angelakis_kay07}.

When the coupling between the cavity field and the two-level system
(which we will just call atom henceforth, noting that they need not
necessarily be only atoms) is very strong (in the so called strong
coupling regime), each cavity-atom unit behaves as a quantum system
whose excitations are combined atom-field excitations called
polaritons. The nonlinearity induced by this coupling or as it is
otherwise known, the photon blockade effect\cite{birnbaum-kimble05},
forces the system to a state where maximum one excitation
(polariton) per site is allowed. However, a superposition of two
different polaritons, which is equivalent to a superposition of two
energy levels of the cavity-atom system, is indeed allowed and
naturally the question arises as to whether that can be used as a
qubit. Purely atomic qubits (formed from purely atomic energy
levels) in cavities have long been discussed in the literature (see
references cited in \cite{alessio}, for example), but such qubits in
distinct cavities do not directly interact with each other unless
mediated through light. On the other hand, a purely photonic field
in a cavity is not easy to manipulate in the sense of one being able
to create arbitrary superpositions of its states by an external
laser. Being a mixed excitation, polaritons interact with each other
as well as permit easy manipulations with external lasers in much
the same manner as one would manipulate and superpose atomic energy
levels. Is there any interesting form of quantum information
processing that can be performed by encoding the quantum information
in a superposition of polaritonic states? While an ultimate aim
might be to accomplish full quantum computation with polaritonic
qubits (it has been recently shown this to possible
 using the cluster state approach \cite{angelakis_kay07}),
we concentrate here on a more modest aim of transferring the state
of a qubit encoded in polaritonic states (a polaritonic qubit) from
one end of the coupled cavity array to another.

\begin{figure}
\centerline{\epsfbox{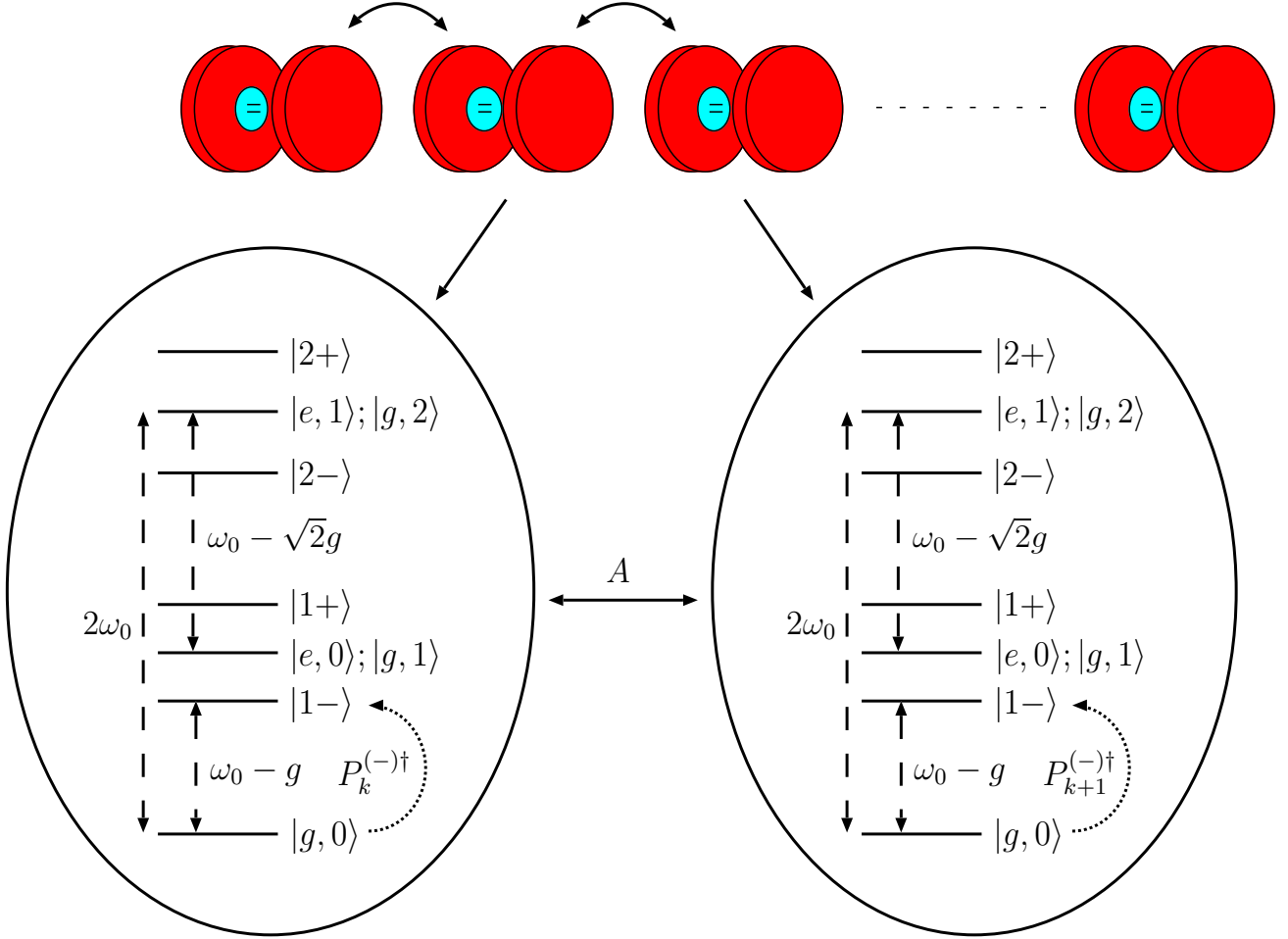}} \caption{A series of coupled
cavities coupled through light and the polaritonic energy levels for
two neighbouring cavities. These polaritons involve an equal mixture
of photonic and atomic excitations and are defined by creation
operators $P_{k}^{(\pm,n)\dagger}=(|g,n\rangle_k\langle g,0|_k\pm
|e,n-1\rangle_k\langle g,0|_k)/\sqrt2$, where $|n\rangle_k,
|n-1\rangle_k$ and $|0\rangle_k$ denote $n, n-1$ and $0$ photon Fock
states in the $k$th cavity. The polaritons of the $k$th atom-cavity
system are denoted as $|n\pm\rangle_k$ and given by
$|n\pm\rangle_k=(|g,n\rangle_k\pm |e,n-1\rangle_k)/\sqrt2$ with
energies $E^{\pm}_{n}=n\omega_{d}\pm g\sqrt{n}$. %and are also
%eigenstates of the the sum of the photonic and atomic excitations
%operator ${\cal N}_k=a_k^\dagger a_k+|e\rangle\langle e|_k$ with
%eigenvalue $n$.
} \label{pol}
\end{figure}

Assume a chain of $N$ coupled cavities. We will describe the system
dynamics using the operators corresponding to the localized
eigenmodes (Wannier functions), $a^{\dagger}_{k}(a_{k})$. The
Hamiltonian is given by
\begin{equation}
H=\sum_{k=1}^{N}\omega_d
a^{\dagger}_{k}a_{k}+\sum_{k=1}^{N}A(a^{\dagger}_{k}a_{k+1}+H.C.).
\end{equation}
 and corresponds to a series quantum harmonic oscillators
coupled through hopping photons. The photon frequency and hopping
rate is $\omega_{d}$ and $A$ respectively and no nonlinearity is
present yet. Assume now that the cavities are doped with two level
systems (atoms/ quantum dots/superconducting qubits) and
$|g\rangle_{k}$ and $|e\rangle_{k}$ their ground and excited states
at site $k$. The Hamiltonian describing the system is the sum of
three terms. $H^{free}$ the Hamiltonian for the free light and
dopant parts, $H^{int}$ the Hamiltonian describing the internal
coupling of the photon and dopant in a specific cavity and $H^{hop}$
for the light hopping between cavities.
\begin{eqnarray}
H^{free}&&=\omega_{d}\sum_{k=1}^N a_k^\dagger a_k+\omega_{0}\sum_k|e\rangle_{k} \langle e|_{k} \\
H^{int}&&=g \sum_{k=1}^N(a_k^\dagger|g\rangle_{k}\langle e|_{k}+H.C.)\\
H^{hop}&&= A\sum_{k=1}^N(a_k^\dagger a_{k+1} +H.C)
\end{eqnarray}
 where g is the light atom coupling strength. The $H^{free}+H^{int}$ part of the
Hamiltonian can be diagonalized
 in a basis of mixed photonic and atomic excitations, called
{\it polaritons} (Fig. 1). While $|g,0\rangle_k$ is the ground state
of each atom cavity system, the excited eigenstates of the $k$th
cavity-atom system are given by $|n\pm\rangle_k=(|g,n\rangle_k\pm
|e,n-1\rangle_k)/\sqrt2$ with energies $E^{\pm}_{n}=n\omega_{d}\pm
g\sqrt{n}$. One can then define polariton creation operators
$P_{k}^{(\pm,n)\dagger}$ by the action
$P_{k}^{(\pm,n)\dagger}|g,0\rangle_k=|n\pm\rangle_k$. As we have
proved elsewhere, due to the blockade effect, once a site is excited
to $|1-\rangle$ or $|1+\rangle$, no further excitation is
possible\cite{angelakis-bose06b}. In simplified terms, this  is
because it costs more energy to add another excitation in already
filled site so the system prefers to deposit it if possible to an a
nearby empty site. This effect has recently lead to the prediction
of a Mott phase for polaritons in coupled cavity
systems\cite{angelakis-bose06b}. If we restrict to the low energy
dynamics of the system such that states with $n\geq 1$ are not
occupied, which can be ensured through appropriate initial
conditions, the Hamiltonian in becomes (in the interaction picture):

%\begin{eqnarray}
%H_{k}^{free}=\omega_{d}\sum_{k=1}^{N}&&P^{(+)\dagger}_{k}P^{(+)}_{k}+
%P^{(-)\dagger}_{k}P^{(-)}_{k}\\
%H_{k}^{int}=g\sum_{k=1}^{N}&&P^{(+)\dagger}_{k}P^{(+)}_{k}-P^{(-)\dagger}_{k}P^{(-)}_{k}\\
%H_{k}^{hop.}=A\sum_{k=1}^{N}&&P^{(+)\dagger}_{k}P^{(+)}_{k+1}+P^{(-)\dagger}_{k}P^{(+)}_{k+1}
%+\nonumber\\
%&&P^{(+)\dagger}_{k}P^{(-)}_{k+1}+P^{(-)\dagger}_{k}P^{(-)}_{k+1}+H.C.
%\label{H_hop_pol}
%\end{eqnarray}
%In the rotating wave approximation, Eq. \ref{H_hop_pol} reads (in
%the interaction picture)
\begin{eqnarray}
H_{I}=A\sum_{k=1}^{N}P^{(-)\dagger}_{k}P^{(-)}_{k+1}+
A\sum_{k=1}^{N}P^{(+)\dagger}_{k}P^{(+)}_{k+1}+H.C. \label{hop}
\end{eqnarray}
where $P_{k}^{(\pm)\dagger}=P_{k}^{(\pm,1)\dagger}$ is the
polaritonic operator creating excitations to the first polaritonic
manifold (Fig. 1). In deriving the above, the logic is that the
terms of the type $P^{(-)\dagger}_{k}P^{(+)}_{k+1}$, which
inter-convert between polaritons, are fast rotating and they
vanish\cite{angelakis-bose06b}.

    We are now in a position to outline the basic idea behind the
protocol. A qubit is encoded as a superposition of the polaritonic
states $|1+\rangle$ and $|1-\rangle$ in the first cavity. The
multi-cavity system is then allowed to evolve according to $H_I$. At
the receiving cavity at the other end we then do a measurement
inspired by a dual rail quantum state transfer protocol
\cite{burgarth-bose05} which heralds the perfect reception of the
qubit for one outcome of the measurement, while for the other
outcome of the measurement the process is simply to be repeated once
more after a time delay. Before presenting the scheme in detail, let
us first present a special set of initial conditions under which
$H_I$ describes the dynamics of two identical parallel uncoupled
spin chains.

  Suppose we are restricting our attention
  to a dynamics in which the initial state is obtained by the action of only one of the operators among $P_k^{(+)\dagger}$ and
  $P_{k}^{(-)\dagger}$ on the state $\prod_k|g,0\rangle_k$ which has all the sites in
  the state $|g,0\rangle$. As $P_k^{(-)\dagger}$ does not act
  after $P_k^{(+)\dagger}$ has acted and vice versa, under the above
  restricted initial conditions, the system is going to evolve only
  according to one of the terms in Eq.(\ref{hop}) {\em i.e.,} only according to the first or the second term.
  To be more precise, if we start with a state
  $P_j^{(+)\dagger}\prod_k|g,0\rangle_k$ only the term
  $A\sum_{k=1}^{N}P^{(+)\dagger}_{k}P^{(+)}_{k+1}$ is going to be
  active and cause the time evolution, while if we start with the
  state $P_j^{(-)\dagger}\prod_k|g,0\rangle_k$ only the term
  $A\sum_{k=1}^{N}P^{(-)\dagger}_{k}P^{(-)}_{k+1}$ will be
  responsible for the time evolution. Each of the operators $P^{(+)\dagger}_{k}$ and $P^{(-)\dagger}_{k}$ individually
  have the same algebra as the Pauli operator $\sigma^{+}_k=\sigma^x_k+i\sigma^y_k$, which makes both the parts
  of the Hamiltonian individually equivalent to a $XY$ spin chain with a Hamiltonian
$H_{XY}=A\sum_k(\sigma^x_k\sigma^x_{k+1}+\sigma^y_k\sigma^y_{k+1})$.
The restricted set of initial states mentioned above can be mapped
on to those of two parallel chains of spins labeled as chain I and
chain II respectively. Let $|0\rangle$ and $|1\rangle$ be spin-up
and spin-down states of a spin along the $z$ direction, $|{\bf
0}\rangle^{(I)}|{\bf 0}\rangle^{(II)}$ be a state with all spins of
both chains being in the state $|0\rangle$, $|{\bf
k}\rangle^{(I)}|{\bf 0}\rangle^{(II)}$ represent the state obtained
from $|{\bf 0}\rangle^{(I)}|{\bf 0}\rangle^{(II)}$ by flipping only
the $k$th spin of chain $I$ and $|{\bf 0}\rangle^{(I)}|{\bf
k}\rangle^{(II)}$ represents the state obtained from $|{\bf
0}\rangle^{(I)}|{\bf 0}\rangle^{(II)}$ by flipping only the $k$th
spin of chain $II$. Then, the restricted class of initial conditions
for polaritonic states can be mapped on to states of the parallel
spin chains as
\begin{eqnarray}
|g,0\rangle_1|g,0\rangle_2....|g,0\rangle_N\rightarrow |{\bf
0}\rangle^{I}|{\bf 0}\rangle^{II},\label{map1}\\
|g,0\rangle_1..|g,0\rangle_{k-1}|1+\rangle_k|g,0\rangle_{k+1}..|g,0\rangle_{N}\rightarrow|{\bf
k}\rangle^{(I)}|{\bf 0}\rangle^{II},\\
|g,0\rangle_1..|g,0\rangle_{k-1}|1-\rangle_k|g,0\rangle_{k+1}..|g,0\rangle_{N}\rightarrow|{\bf
0}\rangle^{I}|{\bf
   k}\rangle^{(II)}
   \label{map3}
\end{eqnarray}
Under the above mapping and under the above restrictions on state
space, $H_I$ becomes equivalent to
   the Hamiltonian of two identical parallel XY spin chains
   completely decoupled from each other. Precisely such a
   Hamiltonian is known to permit a heralded perfect quantum state
   transfer from one end of a pair of parallel spin chains to the other \cite{burgarth-bose05}, and we discuss that below.

     Spin chains are capable to transmitting quantum
states by natural time evolution \cite{bose}. However it is well
known that due to the disperion on the chain \cite{Osborne} the
fidelity of transfer is quite low except for specific engineered
couplings in the spin chains \cite{Christandl,Plenio} or when the
receiver has access to a significant memory
\cite{giovannettiburgath06}. The advantage of the polariton system
is that we have \emph{two parallel and identical} chains. We have
recently shown how this can be made use of in a dual rail protocol
\cite{burgarth-bose05}. The main idea of this protocol is to encode
the state in a symmetric way on both chains. The sender Alice
encodes a qubit $\alpha|0\rangle+\beta|1\rangle$ to be transmitted
as
\begin{equation}
|\Phi(0)\rangle=\alpha |{\bf 0}\rangle^{(I)}|{\bf 1}\rangle^{(II)}
+\beta |{\bf 1}\rangle^{(I)}|{\bf 0}\rangle^{(II)},
\end{equation}
which evolves with time as
\begin{equation}
|\Phi(t)\rangle=\sum_{j=1}^Nf_{1j}(t)(\alpha |{\bf 0}\rangle^{(I)}
|{\bf j}\rangle^{(II)} +\beta |{\bf j}\rangle^{(I)}|{\bf
0}\rangle^{(II)}), \label{phit}
\end{equation}
where $f_{1j}$ is the transition amplitude of a spin flip from the
$1$st to the $j$th site of a chain. Clearly, if after waiting a
while Bob performs a joint parity measurement on the two spins at
his (receiving) end of the chain and the parity is found to be
``odd", then the state of the whole system will be projected to
$\alpha |{\bf 0}\rangle^{(I)} |{\bf N}\rangle^{(II)} +\beta |{\bf
N}\rangle^{(I)}|{\bf 0}\rangle^{(II)}$, which implies the perfect
reception of Alice's state (albeit encoded in two qubits now). The
protocol presented in Ref.\cite{burgarth-bose05} in fact suggested
the use of a two qubit quantum gate at Bob's end which measured both
the parity as well as mapped the state to a single qubit state.
However, here the presentation as above suffices for what follows.
Physically, this protocol, which is called the dual rail protocol,
allows one to perform measurements on the chain that monitor the
location of the quantum information \emph{without perturbing it}. As
such it can also be used for arbitrary graphs of spins (as long as
there are two identical parallel graphs) with the receiver at any
node of the graph. Furthermore, for the Hamiltonian at hand (XY spin
model) it is known \cite{multirail} that the probability of success
converges exponentially fast to one if the receiver performs regular
measurements. The time it takes to reach a transfer fidelity $F$
scales as
\begin{equation}
t=0.33 A^{-1} N^{5/3} |\ln (1-F) |.
\end{equation}

The difference between our current coupled cavity system and the
spin chain system considered in \cite{burgarth-bose05} is that in
our case, the two chains are effectively realized in \emph{one}
system. Therefore, it is not necessary to perform a two-qubit
measurement such as a parity measurement at the receiving ends of
the chain. The qubit to be transferred is encoded as
$\alpha^{'}|1+\rangle_1+\beta^{'}|1-\rangle_1\equiv\alpha|e,0\rangle_1+\beta|g,1\rangle_1$.
This state can be created by the sender Alice using a resonant
Jaynes-Cummings
interaction between the atom and the cavity field. Then the whole
evolution will exactly be as in Eq.(\ref{phit}) with the spin
chain states have to be replaced by polaritonic states according
   to the mapping given in Eqs.(\ref{map1})-(\ref{map3}). The measurement to herald the arrival
   of the state at the receiving end is accomplished by a exciting
   (shelving)
   $|g,0\rangle$ repeatedly to a metastable state by an appropriate
   laser (which does not do anything if the atom is either in
   $|1\pm\rangle$). The fluorescence emitted on decay of the atom
   from this metastable state to $|g,0\rangle$ implies that another
   measurement has to be done after waiting a while. No fluorescence
   implies success and completion of the perfect transfer of the
   polaritonic qubit. Interestingly enough, the measurement at the
   receiving cavity need not be snapshot measurements at regular
   time intervals, but can also be continuous measurements under
   which the scheme can have very similar behavior to the case with
   snap-shot measurements for appropriate strength of the continuous measurement process \cite{kurt}.

 We now briefly discuss the parameter regime needed for the scheme
 of this paper. In order to achieve the required limit of
no more than one excitation per site,  the parameters should have
the following values\cite{angelakis-bose06b}. The ratio between the
internal atom-photon coupling and the hopping of photons down the
chain should be $g/A=10^{2}$. We should be on resonance, $\Delta=0$,
and the cavity/atomic frequencies $\omega_d,\omega_0 \sim 10^4g$
which means we should be well in the strong coupling regime. The
losses should also be small, $g/max(\kappa,\gamma)\sim 10^3$, where
$\kappa$ and $\gamma$ are cavity and atom/other qubit decay rates.
These values are expected to be feasible in both toroidal
microcavity systems with atoms  and stripline microwave resonators
coupled to superconducting qubits \cite{coupled_twolevel}, so that
the above states are essentially unaffected by decay for a time
$10/A$ ($10$ns for the toroidal case and $100$ns for microwave
stripline resonators type of implementations).

  We conclude with a brief discussion about the positive features of
  the scheme and situations in which the scheme might be practically relevant.
The scheme combines the best aspects of both atomic and photonic
qubits as far as communication is concerned. The atomic content of
the polaritonic state enables the manipulation to create the initial
state and measure the received state of the cavity-atom systems with
external laser fields, while the photonic component enables its
hopping from cavity to cavity thereby enabling transfer. Unlike
quantum communication schemes where an atomic qubit first has to be
mapped to the photonic state in the transmitting cavity and be
mapped back to an atomic state in the receiving cavity by external
lasers, here the polaritonic qubit simply has to be created. Once
created, it will hop by itself though the array of cavities without
the need of further external control or manipulation.

  In what situations might such a scheme have some practical
utility? One case is when Alice ``knows" the quantum state she
  has to transmit to Bob. She can easily prepare it as a polaritonic
  state in her cavity and then let Bob receive it through the
  natural hopping of the polaritons. Another situation is when a
  multiple number of cavities are connected with each other through
  an arbitrary graph. The protocol of Ref.\cite{burgarth-bose05}
  still works fine in this situation with Alice's qubit being
  receivable in any of the cavities simply by doing the receiving
  fluorescence measurements in that cavity.

We acknowledge the hospitality of Quantum Information group in NUS Singapore,
and the Kavli Institute for Theoretical Physics where discussions
between DA and SB took place during joint visits. This work was
supported in part by the QIP IRC (GR/S82176/01), the European
Union through the Integrated Projects QAP (IST-3-015848),
SCALA (CT-015714) and SECOQC., and an Advanced Research Fellowship from
EPSRC.

\end{document}